\documentclass{article}
\pdfoutput=1
\usepackage{amsmath}
\usepackage{enumerate}
\usepackage{amssymb}
\usepackage{graphicx}
\usepackage{multirow}
\usepackage{xcolor}
\usepackage{bm}
\usepackage[normalem]{ulem}

\usepackage{jcappub}

\definecolor{darkred}{RGB}{175,0,0}

\def\m@th{\mathsurround=0pt }
\def\eqalign#1{\null\,\vcenter{\openup1\jot \m@th
 \ialign{\strut\hfil$\displaystyle{##}$&$\displaystyle{{}##}$\hfil
 \crcr#1\crcr}}\,}

\begin{document}

%\title{Constraining perturbations beyond the horizon  independently of the cosmological model
%\LV{Constraining spatial curvature independently of the cosmological model with standard sirens and redshift drift// 
\title{Peering beyond the horizon with  standard sirens and redshift drift.}
\author[1,2]{Raul Jimenez,}
\author[1]{Alvise Raccanelli$^\star$\let\thefootnote\relax\footnote{$^\star$Marie Sk\l{}odowska-Curie fellow},}
\author[1,2]{Licia Verde,}
\author[3,4,5,6]{Sabino Matarrese}

\affiliation[1]{ICCUB, University of Barcelona, Marti i Franques 1, Barcelona, 08028, Spain}
\affiliation[2]{ICREA, Pg. Lluis Companys 23, Barcelona, 08010, Spain} 
\affiliation[3]{Dipartimento di Fisica e Astronomia ``G. Galilei'', Universita'  degli Studi di Padova, via Marzolo 8, I-35131, Padova, Italy}
\affiliation[4]{INFN, Sezione di Padova, via Marzolo 8, I-35131, Padova, Italy}
\affiliation[5]{INAF-Osservatorio Astronomico di Padova, Vicolo dell'Osservatorio 5, I-35122 Padova, Italy}
\affiliation[6]{Gran Sasso Science Institute, INFN, Viale F. Crispi 7, I-67100 L'Aquila, Italy}

\emailAdd{raul.jimenez@icc.ub.edu; alvise@icc.ub.edu; liciaverde@icc.ub.edu}

\abstract{An interesting test on the nature of the Universe is to measure the global spatial curvature of the metric in a model independent way, at a level of $|\Omega_k|<10^{-4}$,  or, if possible, at the cosmic variance level of  the amplitude of the CMB fluctuations $|\Omega_k|\approx10^{-5}$.
A limit of $|\Omega_k|<10^{-4}$ would yield stringent tests on several models of inflation. Further, improving the constraint by an order of magnitude would help in reducing ``model confusion" in standard parameter estimation. Moreover, if the curvature is measured to be at the  value of the amplitude of the CMB fluctuations, it
would offer a powerful test on the  inflationary paradigm and
would indicate that our Universe must be significantly larger than the current horizon.
On the contrary, in the context of standard inflation, measuring a value above CMB fluctuations will lead us to conclude that the Universe is not much larger than the current observed horizon; this can also be interpreted as the presence of large fluctuations  outside the horizon. However, it has proven difficult, so far, to find observables that can achieve such level of accuracy, and, most of all, be model-independent. Here we propose a  method that  can in principle achieve that; this is done by making minimal assumptions and using  distance probes that are cosmology-independent: gravitational waves, redshift drift and cosmic chronometers.  We discuss what kind of observations are needed  in principle to achieve the desired accuracy.} 

\maketitle

\section{Introduction} 

While significant progress --e.g.,~\cite{Planck15}-- has been made in  measuring some of the parameters of the standard model of cosmology ($\Lambda$CDM), we still lack a physical understanding of what these parameters mean. For example, we have no clue about what the major energy components of the Universe are, namely: dark matter and dark energy. The situation is even worse as we do not even understand within the context of the standard model of particle physics why there is matter at all and not only photons; this is the so-called matter-antimatter asymmetry. Clearly, progress needs to be made to understand  the physics of the Universe going beyond simple parameter fitting within a given model.

One route of making progress is to measure properties of the Universe in a model-independent way, as the majority of the current parameter determination from the cosmic microwave background (CMB) and large scale structure (LSS) data are assuming a cosmological model, which relies on a suite of assumptions. On the theoretical front, there is significant interest in modeling the metric of the Universe without the constraints of imposing homogeneity and isotropy. One of the pioneering approaches has been the ``silent Universes" approach~\cite{Matarrese:1994, Bruni:1995}. 
On the experimental side, one attempt was the determination of the standard ruler in a model independent way~\cite{ruler,ruler1}. In this work, we take this approach one step further and develop a method to measure the spatial curvature of the Universe in a  way that is independent of the cosmological model. 

While  current measurements have established that curvature is dynamically negligible, precise limits, if possible relying on minimal cosmological assumptions, will  enable a host of powerful tests of cosmology. 
Measuring the spatial curvature of the Universe, $\Omega_k$,  is a very important challenge of current observational cosmology (e.g., for constraining inflationary models~\cite{Linde:2003, Linde:2014, Guth:2012, Guth:2014}).

 Precision measurements of spatial curvature can be used to test the cosmological principle of homogeneity and isotropy itself~\cite{Clarkson}.
There has been some recent work (see e.g.,~\cite{Vardanyan:2009, Vardanyan:2011, Takada:2015}) proposing model-dependent methods to measure the curvature and some methods to reduce the measurement sensitivity to dark energy~\cite{Knox,Witzemann:2017lhi}. 
Measurements of the curvature parameter come from the combination of CMB~\cite{Kamionkowski:1994, Jungman:1996} and LSS~\cite{Takada:2015, Bull, DiDio:2016}. Current limits on spatial curvature from observations from the Planck satellite after adding BAO data have established an upper limit on its value of $|\Omega_k| < 5 \times 10^{-3}$~\cite{Planck15}, while the combination of Planck and the BOSS survey data found $\Omega_k = 0.0010\pm0.0029$~\cite{Sanchez:2014}.
These constraints are all  obtained within the framework of the $\Lambda$CDM model (see Ref.~\cite{Ratra} for constraints with other cosmological models yielding to a detection of curvature).

It is important to note that, based on statistical arguments, the information coming from joint CMB and future large-scale structure surveys, may not allow unambiguous conclusions about cosmology  and  the geometry of our Universe if the value of the curvature parameter is  at or below $|\Omega_k|\sim\mathcal{O}(10^{-4})$~\cite{ClarksonCortes, Vardanyan:2009, Licia, Vardanyan:2011}. Simply fitting the curvature as a parameter in a model-dependent way results in ``model confusion" i.e., a Bayesian analysis would wrongly favour a flat Universe in the presence of curvature for models that deviate from the standard $\Lambda$CDM. It is this model-dependence of the curvature determination that motivates  measuring the curvature of the Universe at  the level of primordial curvature perturbations of $\sim 10^{-5}$ e.g., Ref.~\cite{Waterhouse:2008}. The cosmology-independent determination that we advocate here  addresses this issue in a complementary and synergistic way.  

There are clearly different  aims in  constraining curvature. 
One is   to clearly  improve the current $\Lambda$CDM-based determinations of $\Omega_k$ below its current value of $\sim 5 \times 10^{-3}$ using standard, cosmological-model-dependent, approaches. This will not be addressed here. We focus here on  obtaining constraints in a cosmological model-independent way. In this context there are several target levels for the constraints:  confirming  the current bounds, one or two orders of magnitude improvements over current constraints i.e., $|\Omega_k|<10^{-4}$, and the cosmic variance floor of $|\Omega_k|<10^{-5}$. Constraints at the  $|\Omega_k|<10^{-4}$ will offer stringent tests on several models of inflation;
while inflation generally predicts the spatial curvature of the Universe to be essentially flat, so that detecting $\Omega_k\neq 0$ would significantly limit the parameter space for inflationary models,
scenarios involving interesting open~\citep[see e.g.,][]{Gott:1982, Kamionkowski:1994a, Kamionkowski:1994b, Linde:1995, Bucher:1995, Linde:1998, Freivogel:2006} or closed~\cite{Linde:2003} geometries have been considered.
Moreover, a negative curvature arises in false vacuum decay, which can naturally arise in the multiverse~\cite{Freivogel:2006}, while this would require strong fine tuning in standard cosmological models. Conversely, a measurement of positive curvature would falsify most of the multiverse models.

Constraints at the $|\Omega_k|<10^{-5}$ levels will achieve two goals: eliminate model confusion, and offer a powerful test of the inflationary paradigm. If inflation is the correct model to describe the early Universe, then our current observed horizon is just a small patch of a very large (infinite) Universe, hence  curvature of the metric of the Universe should be at the level of the fluctuations which we know experimentally have a value of $\delta \rho / \rho \sim 10^{-5}$.

There are two main complications to achieve our goal: first, it is difficult to find observables that are independent of the cosmological model and second, it is difficult to devise a method to measure the curvature at the required precision; Ref.~\cite{Bull} shows the difficulties of  obtaining measurements of the curvature that approach the intrinsic limit of  $|\Omega_k| < 10^{-5}$ even  within a $\Lambda$CDM model, and in~\cite{DiDio:2016} it was shown that adding large-scale relativistic effects will not help enough. This paper attempts to take up this challenge and evaluate what kind of observations will be needed to achieve this precision.

As arduous as it may be, a model-independent determination of the value of the curvature will provide a powerful consistency test and would shed light on  early Universe physics and what may lie beyond our current horizon.

\section{Method}
We start by only assuming homogeneity and isotropy and hence a Friedman-Robertson-Walker (FRW) metric; only when needed we will explicitly point out when we assume the validity of Einstein's General Relativity (GR). Our considerations and calculations are valid in any metric theory of gravity, as long as homogeneity and isotropy hold, and can be easily generalized to departures from GR.
Some considerations are in order here.  If the geometry of our Universe slightly deviates from the FLRW geometry, then  definition of global spatial curvature  in the FLRW sense does not necessarily  hold. One case explored in the literature is for example that of emerging spatial curvature \cite{BuchertCarfora2008,Royetal.2011,2017arXiv170701800B,2017arXiv171202967B,2018CQGra..35b4003B}, where  it is the very same non-linear evolution of cosmic structures what causes deviations from FLRW metric.
Since EinsteinÕs equations are non-linear, the evolution of an inhomogeneous nonlinear system homogeneous and isotropic  only in a statistical sense is not the same as the evolution of an exactly homogeneous and isotropic system  \cite{Buchertetal2015}. A detailed quantitative understanding of this phenomenon  requires fully relativistic cosmological simulations. It has been shown \cite{2017arXiv171202967B} that  once the  evolution of structures enters the non-linear regime, the symmetry between overdensities and underdensities is broken. Then one can still measure an effective  mean spatial curvature of the universe, but this one  slowly drifts  towards negative curvature induced by cosmic voids.  The emergence of the spatial curvature (due to nonlinear relativistic corrections) then biases the determination of a global curvature parameter.
 For our application, the measurement of curvature parameter,  if then reinterpreted  in a FLRW metric, will be biased. It will have to be interpreted including  relativistic corrections due to nonlinear cosmic evolution.  Keeping all the in mind, we proceed assuming a FRW metric.

The luminosity distance in a FRW metric, as a function of redshift $z$ and the Hubble constant, $H_0$, and Hubble parameter, $H(z)$, is:
\begin{equation}
d_L(z)=\frac{(1+z)}{H_0\sqrt{|\Omega_k|}}S_k\left[\sqrt{|\Omega_k|} \int_0^z\frac{H_0}{H(z')}dz'\right] \, ,
\label{eq:dl}
\end{equation}
where $S_k(x)$ is a function that is $\sin(x)$ if $\Omega_k<0$, $\sinh(x)$ if $\Omega_k>0$ and $x$ if $\Omega_k=0$.
It is therefore clear that the combination of  observational measurements of $d_L(z)$, $H_0$ and $H(z)$ will provide an estimate for $\Omega_k$ (see \cite{ruler,ruler1})\footnote{While $\Omega_k$ is usually interpreted in the context of GR, in this more general context $\Omega_k$ stands for $k c/(H_0 R_0)^2$ where: $k$ determines the curvature of the Universe and can take only values $-1,0,1$ and $R_0$ is the present value of the scale factor; $c$ is the speed of light.}.
However, the complication is to obtain a measurement of $\Omega_k$ at the required  level of precision. 
This is discussed later in more detail. Ref.~\cite{Clarkson} also starts from eq.~\ref{eq:dl} above to test instead the cosmological principle, arguing that in a FRW metric $\Omega_k$ should be constant and does not depend on redshift. They proceed then by taking  redshift derivatives of $d_L$ and $H(z)$ to detect deviations from the null hypothesis, and use standard cosmological-model dependent observable quantities as data. Here we proceed differently.  We assume the FRW metric.
We wish to measure the expansion history and luminosity distances using quantities that are not dependent on the cosmological model themselves. The newest development is that it has been recently experimentally demonstrated that a gravitational wave detection with an electromagnetic counterpart will provide $z$ and $d_L(z)$ independently of the cosmological model~\cite{Schutz, Abbott:2017a, Abbott:2017b, Abbott:2017c}. This technique has been named ``standard sirens".

\subsection{Observables}

We consider two routes for measuring a combination of $H_0$ and $H(z)$ with high accuracy: cosmic chronometers (CC) and redshift drift, as supernovae have been discussed elsewhere~\cite{Bull}. For $d_L$ measurements we will consider standard sirens.

Redshift drift $\Delta z$, and  velocity shift  $\Delta v$, over a time interval $\Delta t_0$ for the observer, are related to the Hubble expansion as
\begin{equation}
\frac{\Delta v}{\Delta t_0}=\frac{\Delta z}{\Delta t_0}\frac{1}{1+z}=H_0-\frac{H(z)}{1+z}
\end{equation}
where $\Delta t_0$ is usually of the order of tens of  years.
With an estimation of the uncertainty on $\Delta v$  over several redshift bins from future surveys we can calculate all that is needed to obtain some forecasts on uncertainties on $H_0$ and $H(z)$. Estimates of uncertainties in future redshift drift measurements can be found in 
e.g.~\cite{Liske:2008, Martins:2016} which suggest the following formula for the error in velocity drift (in cm/s):

\begin{equation}
\sigma(\Delta v)= \frac{1.35}{6200} \frac{2370}{SNR}\sqrt{\frac{30}{N_{QSO}}}\left(\frac{5}{1+z}\right)^{\alpha(z)};
\end{equation}
where $N_{QSO}$ is the number of observed quasars, and $\alpha(z)=1.7$ for $z<4; \alpha(z)= 0.9$ beyond that redshift. 
Taking as an example observations with the E-ELT, in a time span of 30 years we can assume SNR=3000 for 10 QSOs for each redshift bin in Table~\ref{table:1}.

The other alternative to obtain $H_0$ and $H(z)$  is to use cosmic chronometers \cite{clocks,Simonetal, moresco1,moresco2}. They measure directly 
\begin{equation}
H(z) = \frac{-1}{1+z} \frac{dz}{dt} \; .
\end{equation}
Significant work has been done to assess the feasibility of the method and current measurements of $H(z)$ are at the 6\% level \cite{moresco1,moresco2} but are limited by systematic errors. Future measurements from EUCLID \cite{moresco2}  will be instrumental to reduce systematic errors, and will easily achieve sub-percentage  statistical precision; further improvements could be achieved with even larger surveys. 
 
The advantage of the redshift drift method is that it does not depend on any modelling of the properties of QSOs, while the cosmic chronometer method does depend on modelling of stellar evolution (they are basically atomic clocks). On the other hand, the cosmic chronometer method can obtain large (millions) samples of stellar clocks, thus reducing the statistical uncertainties on $H(z)$ significantly. 

In what follows, we consider both the redshift drift error and a  1\% error on $H(z)$ in 8 redshift bins up to $z=5$, see Table~\ref{table:1}. Then we will compute what error on $H(z)$ and $d_L(z)$ is needed to achieve a target error on $\Omega_k$. 

In our proposed approach, luminosity distance  estimates and their errors come from gravitational waves (GWs) measurements with optical counterpart. The recent discovery of the neutron star-neutron star merger by the LIGO-VIRGO collaboration~\cite{Abbott:2017a, Abbott:2017b, Abbott:2017c}, shows that it is feasible to measure the luminosity distance $d_L$  even for a single object with error-bars highly competitive compared to e.g.,  the standard candles (supernovae) approach.  We propose to combine two measurements, also known as standard sirens approach~\cite{Schutz}, to measure the luminosity distance $d_L$ from gravitational waves, and the redshift from the optical counterpart (we will assume GWs from mergers with EM counterpart).

Even with planned GW detectors, it will be possible to measure $d_L(z)$ at 1\%~\cite{Cutler:2009}. Moreover, lensed GWs have been shown to be even more powerful, allowing a 0.7\% measurement of $H_0$ with only 10 objects~\cite{Liao:2017} (but assuming flatness). 

To measure $d_L$  using GW we will assume that we will have 8 redshift bins in the range $0 < z < 5$.
For this purpose we assume a futuristic instrumental setup based on the Big Bang Observer~\cite{BBO}; errors on $d_L$ are given in Table~\ref{table:1} and,  given the current uncertainties in key quantities such as  the experimental set up and expected number of sources,  need to be taken as an order of magnitude  estimate. In a futuristic outlook, they could even be improved (see e.g.,~\cite{Cutler:2009}).

\subsection{Observational issues}
\label{sec:obs_issues}
There are, however, three complications: {\it a)} local peculiar velocities  (also called redshift space distortions, RSD~\cite{Kaiser:1987}); {\it b)} gravitational lensing~\cite{Dai:2017a, Dai:2017b}; {\it c)} cosmological perturbations that modify the GW wavefront, giving rise to an additional uncertainty on the determination of $d_L$~\cite{Bertacca:2017} (these include integrated Sachs-Wolfe and doppler effects). Regarding point {\it (a)}, on average, there will be a $\approx$ few $\times 100$ km/s peculiar velocity effect; on large scale these are bulk flows generated by linear RSD -a.k.a. Kaiser effect~\cite{Kaiser:1987}-, while on small scales  peculiar velocities are local and randomized in direction -a.k.a. fingers-of-God. The effect of large-scale flows could in principle be reduced or removed by having a template of the density field and assuming GR to predict from it the peculiar velocity field. It will also average out if enough independent ``patches" are observed. The random  component   is expected  to average out to zero in the case where a large number of GWs are detected, especially if not too nearby and  over large patches in the sky (for a review see e.g.,~\cite{KaiserHudson} and references therein).  A linear order estimate indicates a relative  distance error of $\sim 10^{-3}$ per object hence, especially at small scales it quickly averages out. This effect is common to Supernovae determination of luminosity distances, but it is highly subdominant to other effects that contribute to the scatter of the Supernovae-based luminosity-distance relation. We will therefore neglect this contribution in the following. Effects of peculiar velocities on the template of the GW signal were discussed e.g., in~\cite{Bonvin}. We assume these  effects can also be corrected for.

We refer to the effect of perturbations described in points {\it (b)} and {\it (c)} as ``projection effects''. In Table~\ref{table:1} we report also the expected relative error per redshift bin due to these effects, following Ref.~\cite{Bertacca:2017}. The magnitude of this effect is highly uncertain because the GW equivalent of magnification bias is poorly known. We assume that this error also behaves as a statistical error rather than a systematic one. In other words, it does not yield a systematic error-floor but an extra source of scatter that gets reduced by averaging the signal from sources widely separated in the sky. Thus this error will be summed to the standard observational one which we refer to as ``instrument''. It is important to note that in the redshift range of interest the dominant term in the projection effects is the one due to lensing convergence (see~\cite{Bertacca:2017}). The magnitude of this contribution depends critically on the equivalent luminosity function of the GW signal, in analogy of the effect of cosmic magnification in optical wavelength that depends crucially on the slope of the source galaxy luminosity function.
In Table~\ref{table:1} we report the values of projection additional errors calculated when the GW-equivalent magnification bias is set to be $s=0$ (it is useful to recollect that $s=0.4$ would cancel lensing convergence effects), therefore giving quite pessimistic results.
There are in principle two ways to minimize this source of errors: {\it i)} by selecting a sub-sample of the sources (as suggested in~\cite{Raccanelli:Doppler}) so that the source luminosity function has a magnification bias that cancels convergence effects (as the velocity and ISW-like contributions are almost negligible), or
{\it ii)} by performing a ``de-projection'' of the observed map, in the same way CMB maps are now routinely de-lensed~\cite{Kesden:2002, Hirata:2003, Seljak:2004, Hanson:2010}, having a 3D galaxy density map and a bias estimate available. In the same spirit, it is in principle possible to ``de-RSD'' our observed map, hence being able to completely correct for local and perturbations modifications of the estimate of $d_L$. 

For this reason, in the following we will consider two cases, one where the projection effect is fully present and one where it is subdominant compared to the statistical (instrument) errors and therefore is neglected (see Table~\ref{table:1}). These should be considered as two (pessimistic and optimistic) limiting cases.

\section{Forecasts}

In this section we first forecast the achievable error on $\Omega_k$ given the experimental set up of  Table~\ref{table:1}. We then  investigate what kind of observations and what level of uncertainty  are needed to measure the curvature at the cosmic variance level of  $10^{-5}$ with the method proposed in the previous section.  We finally reflect on whether this is achievable  at least in principle or if there is a fundamental limitation that prevents it.

In interpreting our results one should keep in mind that for this analytical forecast the  additional ``scatterÓ on the determination of $d_L$ due to lensing magnification has been modelled as Gaussian. In reality it is well known that   the distribution is non-gaussian but its shape can be well approximated by a known analytic formula see e.g., \cite{1106.3823}. In any practical application this will have to be correctly modelled to avoid important biases. With this caveat in mind, for this initial error-estimate we  still use the Gaussian approximation.

\subsection{Analytic considerations}
\begin{figure}[h]
 \centering
 \includegraphics[width=0.7\textwidth]{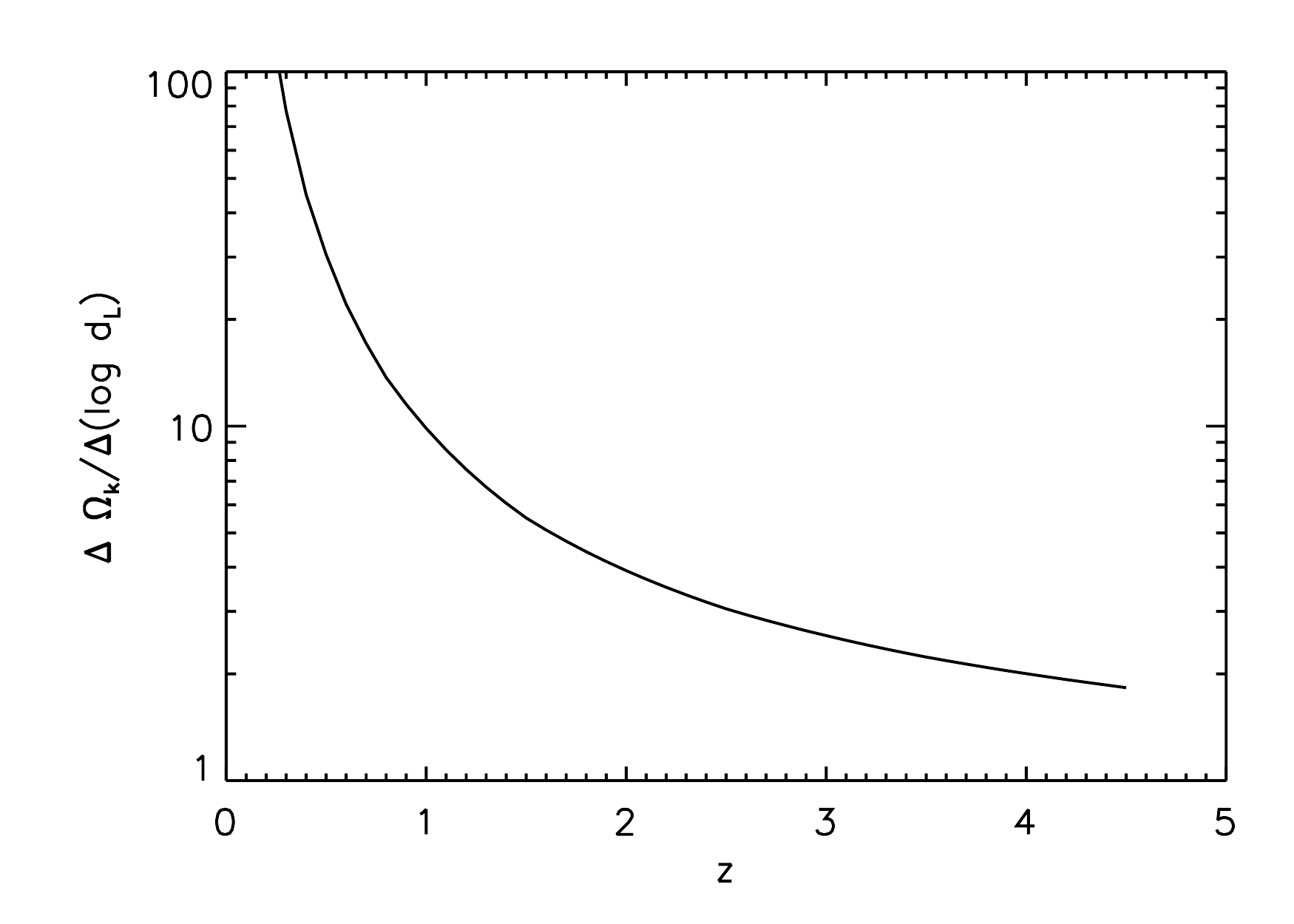}
 \caption{Response of the inferred  $\Omega_k$ values to a variation in $\Delta d_L/d_L\equiv \Delta \log d_L$ as a function of redshift (assuming that the Universe expansion history is fixed and known with negligible errors). This has been computed using Eq.~(\ref{eq:analytic}) for a fiducial concordance $\Lambda$CDM cosmology.}
 \label{fig:analytic}
 \end{figure}

It is worth noting that only the luminosity distance has information on geometry and thus curvature. Let us define the comoving distance as
$\chi(z)=\int_0^z H(z')^{-1}dz'$. With this definition the luminosity distance is related to $\chi$ by 
\begin{equation}
d_L(z)=(1+z) (\sqrt{|\Omega_k|}H_0)^{-1}S_k(H_0\chi(z)\sqrt{|\Omega_k|})\,.
\end{equation}
 Recall that a Taylor expansion of $\sin(x)$ around zero yields $\sin(x)\sim x-x^3/6$ and for $\sinh(x)$ we have  $\sinh(x)\sim x+x^3/6$. 
Hence we obtain that to leading order 
\begin{equation}
d_L(z)=(1+z)\left[\chi(z)+\frac{\Omega_k}{6}H_0^2\chi(z)^3\right]\,.
\label{eq:analytic}
\end{equation}
Therefore, assuming negligible uncertainty on  $\chi(z)$, we can evaluate the variation of  $d_L$ due to a variation on
$\Omega_k$  and viceversa. Figure~(\ref{fig:analytic}) shows the variation on $\Omega_k$ per relative variation  on $d_L$ as a function of redshift, or to be more explicit, $\left(\partial \log d_L/\partial \Omega_k\right)^{-1}$ computed from the expression in Eq.~(\ref{eq:analytic}).  It is easy to see why it is so challenging to reduce the error on the curvature below $~10^{-3}$: even assuming a perfect reconstruction of the expansion history -- $H(z)$, and hence $\chi(z)$ --, the error on    
$\Omega_k$  in the redshift range of interest is always 
larger than that on $d_L$.  But while $d_L$ changes with redshift, $\Omega_k$ does not. This indicates that many high-precision measurements of $d_L$ at different redshifts are needed and that, for the same precision on $d_L$,  higher $z$  are better suited to measure $\Omega_k$.

As it can be seen from Figure~(\ref{fig:analytic}), we need a relative measurement of $d_L$ with precision better than $10^{-4}$ to obtain constraints on the curvature better than $10^{-3}$ at low redshift, while at high redshift   the ratio between these two errors tends to $\sim 2$. As discussed in Section~2.2, while at high redshift one is expected to find more sources and thus reduce the $d_L$ ``instrumental" error, the scatter on $d_L$ due to projection effects increases.  While we have argued that projection effects could at least in principle be removed (or made subdominant) either at the expenses of reducing the number of sources (to tailor the sample to have a magnification bias that cables the effect), or by a reconstruction effort, this consideration indicates that there may be a trade-off between  target redshift  range and  number of sources  that may minimise the combination of the instrumental plus projection effects errors optimising observational efforts. This  will be discussed elsewhere.

\subsection{Uncertainty estimates}

We use a Monte Carlo numerical method to invert Eq.~\eqref{eq:dl} to compute the expected errors on $\Omega_k$, when using the errors on the observables in Table~\ref{table:1}.  We also include a prior on $H_0$ with the reported error as in Table~\ref{table:2}. As discussed above, it is possible to use either cosmic chronometers or redshift drift to obtain a cosmological model independent measurement of $H(z)$. The results are reported in Table~\ref{table:2}, including or not projection effects, as explained in Section~\ref{sec:obs_issues}.  Table~\ref{table:2} shows that errors of the order $\sigma_{\Omega_k}\sim 10^{-2}$ can be achieved with a set up similar to that of Table~\ref{table:1}. The error on the curvature is dominated by the  uncertainties on $d_L$. Moderate improvements on $d_L$ (an order of magnitude)  are required to reach the level of $\sigma_{\Omega_k}\sim 10^{-3}$.  To reach the level $\sigma_{\Omega_k}\sim 10^{-4}$, two orders of magnitude improvements on distance measurements are required.

\begin{table}
\caption{Uncertainties as a function of redshift for the different observables used to obtain constraints on $\Omega_k$ using the cosmology-independent method described in the text. See text for more details about the different survey assumptions.}
\begin{center}
\begin{tabular}{ccccc}
\hline
& Instrument& Projection& & cm/s  \\
$z$ & $\sigma_{d_L}^I/d_L$& $\sigma_{d_L}^P/d_L$ & $\sigma_{H(z)}/H(z)$ &$\sigma_{\delta v}$  \\
\hline
0.2  &   0.001 & 0.002& 0.01 &0.0034  \\
0.4   &  0.001 &0.003 & 0.01 &0.0026  \\
0.6    &  0.001 & 0.006  & 0.01 &0.002 \\
0.8    &  0.001 &0.007 & 0.01 &0.0017  \\ 
1.5  &   0.002  &0.008& 0.01 &0.00097  \\
2.5  &  0.003  &0.02 &  0.01 &0.00055  \\
3.5  &  0.005   &0.03 & 0.01 &0.00036  \\
4.5  &   0.007  &0.04 &0.01 &0.00025 \\
\hline
\end{tabular}
\end{center}
\label{table:1}
\end{table}%

\begin{table}
%:
\caption{Constraints on $\Omega_k$ for different assumptions on the uncertainties of the observables as in Table~\ref{table:1}. We explore what reduction of  the uncertainty on $d_L$ and expansion history are needed to achieve constraints at the level of $\Omega_k \sim 10^{-5}$. The $``-"$ in the case with projection indicates that reduction of the other errors is irrelevant since this term already dominates.}
\begin{center}
\begin{tabular}{ccc}
\hline
& No projection& With projection\\
&    $\sigma\Omega_k$& $\sigma\Omega_k$\\
\hline
\hline
redshift drift&&\\
\hline
 2.5\% $\sigma(H_0)$ & 0.017 &0.033\\
1\% $\sigma(H_0)$ & 0.008 &0.026\\
1\% $\sigma(H_0)$ and $\sigma(d_L)/10$    & 0.001 & -- \\
1\% $\sigma(H_0)$ and $\sigma(d_L)/10$ and $\sigma_{\delta v}/10$   & 0.0002 & -- \\
 1\% $\sigma(H_0)$ and $\sigma(d_L)/1000$ and $\sigma_{\delta v}/10$  & $3 \times 10^{-5}$ & -- \\
\hline
\hline
cosmic clocks&&\\
\hline
2.5\% $\sigma(H_0)$& 0.019 & 0.05\\
 1\% $\sigma(H_0)$& 0.008 & 0.02 \\
1\% $\sigma(H_0)$ and $\sigma(d_L)/10$ & $8 \times 10^{-4}$ & -- \\
 1\% $\sigma(H_0)$ and $\sigma(d_L)/100$ and $\sigma_{H(z)}/10$ & $3 \times 10^{-4}$ & -- \\
  1\% $\sigma(H_0)$ and $\sigma(d_L)/1000$ and $\sigma_{H(z)}/10$ & $2 \times 10^{-5}$ & -- \\
\hline
\end{tabular}
\end{center}
\label{table:2}
\end{table}%

We have shown above that planned experiments will be able to provide cosmological model-independent measurements on the spatial curvature at the level $\sigma_{\Omega_k} \sim \mathcal{O}(10^{-2})$. 
A moderate improvement of such experiments could allow reaching levels of $\sigma_{\Omega_k} \sim  10^{-3}$.
This would confirm the current conclusion on the level of spatial flatness coming from  cosmology-model-dependent measurements. A more significant  improvement of such experiments  could allow reaching levels of $\sigma_{\Omega_k} \sim  10^{-4}$, which would represent a considerable advancement in the primordial Universe model testing.

The next step is trying to measure the curvature at  the CMB fluctuations level, therefore obtaining information on global properties of the Universe, in the context of standard inflation. Thus the question becomes: what is needed to obtain $\sigma(\Omega_k) \lesssim 10^{-5}$, comparable to the cosmic variance limit of $\approx 1.5 \times 10^{-5}$ \cite{Waterhouse:2008}? From  Table~\ref{table:2} we can see that, with our assumed setup, one could measure the curvature at the perturbation level only when improving the errors on the luminosity distance and the Hubble parameter by a factor of 1000 and 10, respectively. This might seem unfeasible if we were to simply scale the errors with the square root of the number of sources. However,  improvements on measurements of the orientation of the GW source via multiple GW observatories and on the interferometer noise, will ease the increase on the number of sources needed to reach $\sigma_{\Omega_k} \sim 10^{-5}$, much less than if current noise on GW is assumed, as we have done in Table~\ref{table:2}.

Our results provide a clear target for 
what would be, in principle,  the requirements needed to obtain such measurements. Whether this is achievable in practice will need to be evaluated. In the next, final Section, we will discuss the implications of having such knowledge of the spatial curvature.
The calculation presented here should be considered as a Òproof of principleÓ. The required experimental errors to achieve a cosmic variance level for the curvature parameter correspond to a likely idealised case, even if the LISA mission may reach this level of accuracy. In any case, our results can be used as a guidance to strengthen the science case for future experiments.

\section{Discussions and Conclusions}
In this paper we have proposed a method to obtain measurements of the global spatial curvature of the universe independently of the cosmological model.
We have shown that a combination of gravitational wave observations  (standard sirens) with redshift drift or cosmic chronometers measurements of the expansion history could provide enough accuracy to obtain measurements of  $\Omega_k < 10^{-3}$, with only moderate improvements on the specifications of currently planned experiments.
While the current limit on the curvature provided by the Planck satellite in combination with the BOSS survey is of $\Omega_k < 5 \times 10^{-3}$~\cite{Planck15}, this is obtained within the framework of a cosmological model ($\Lambda$CDM). In contrast, our method is independent of the cosmological model.
Finally, we argue that with a more futuristic setup it is, in principle, possible to reach a precision in the measurement of the spatial curvature that approaches the limit imposed by primordial fluctuations; a measurement of $\Omega_k$ using the proposed method here will therefore provide clues on the size of our Universe and the amount of fluctuations beyond our current horizon, an important and unprobed source of information on the origin of the Universe. \\

One of the most appealing features of measurements of $\Omega_k$ is the fact that the spatial curvature lets information leak in from outside our observable patch; in a sense, it is the window that allows us to peak outside the {\it in principle observable Universe}.

We point out that a measured value of $\Omega_k > 10^{-5}$ would allow us to speculate on the size of the entire Universe, but  only in the context of standard inflationary scenarios: it is in fact possible to imagine situations in which the spatial curvature is large even if the Universe is also considerably larger than the observable patch.
If, e.g., the initial curvature was very high, with a suitable number of e-foldings the Universe can be very large even with a value of $\Omega_k > 10^{-5}$.

Finally, it is interesting to note an often overlooked point: the very concept of cosmic variance is related and due to the ``fair sample hypothesis''~\cite{Peebles:1980}.
This involves ergodicity and the fact that we assume the Universe to be homogeneous and isotropic.

One could be tempted to state that cosmic variance is a byproduct of inflation; however, this is not true, because at its very core, cosmic variance arises from the fact that it is assumed that our observable patch of the Universe is a single realization of some stochastic process~\cite{Neyman:1962}, hence we need to treat our patch as part of a larger ensemble. In fact this idea was introduced in the 1960s, well before inflation was theorized. While it is true that with inflation there are N (i.e., a large number of) realizations that exist in N spatial patches, and we live in one of them (and hence the statistical ensemble), even without inflation, under the fair sample hypothesis, our Universe is one particular realization (even in the purely hypothetical case where there is only one patch and nothing outside), out of N possible Universes that {\it could} have been created -- for an early discussion about primordial fluctuations well before any inflationary models, see Ref.~\cite{Silk:1967}.

However, if the entire Universe (i.e., a patch at least much larger than the {\it currently observable} Universe) is emerging from some deterministic process, or if there is some underlying physical law imposing fixed perturbations (of a certain amplitude), then there is no intrinsic cosmic variance as commonly defined. A careful investigation of which underlying physical models might satisfy the above condition is beyond the scope of this paper, and while we do not necessarily endorse this possibility, given our attempt of being as model-independent as possible, it makes sense to take into account the possibility that, cosmic variance is not an obligatory feature of our Universe\footnote{Note that in this context, sample variance is not the same as cosmic variance, and the first can exist within the local universe even if the second is not present.}. \\

\bigskip
{\it Acknowledgments.}
RJ and LV thank the Radcliffe  Institute for Advanced Study at Harvard University for hospitality. Funding for this work was partially provided by the Spanish MINECO under projects AYA2014-58747-P AEI/FEDER UE and MDM-2014-0369 of ICCUB (Unidad de Excelencia Maria de Maeztu). AR has received funding from the People Programme (Marie Curie Actions) of the European Union H2020 Programme under REA grant agreement number 706896 (COSMOFLAGS). LV acknowledges support of  European Union's Horizon 2020 research and innovation programme ERC (BePreSySe, grant agreement 725327).

\end{document}